\documentstyle[aps,graphicx]{revtex}
\begin{document}

\draft
\twocolumn[\hsize\textwidth\columnwidth\hsize\csname@twocolumnfalse\endcsname

\title{Quantum Phase Fluctuations Responsible for Pseudogap}

\author{V.P.~Gusynin$^1$, V.M.~Loktev$^1$\cite{e-mail},
R.M.~Quick$^2$ and S.G.~Sharapov$^{1,2,3}$\cite{e-mail2}}

\address{
$^1$Bogolyubov Institute for Theoretical Physics, 03143 Kiev, Ukraine\\
$^2$Department of Physics, University of Pretoria, 0002 Pretoria, South Africa\\
$^3$ Institut de Physique, Universite de Neuch\^atel,
CH-2000 Neuch\^atel, Switzerland }

\date{27 December 2000}

\maketitle

\begin{abstract}
The effect of ordering field phase fluctuations on the normal and
superconducting properties of a simple 2D model with a local
four-fermion attraction is studied. Neglecting the coupling
between the spin and charge degrees of freedom an analytical
expression has been obtained for the fermion spectral function as
a single integral over a simple function. 
From this we show that, as the temperature increases through the 2D
critical temperature and a nontrivial damping for a phase
correlator develops, quantum fluctuations fill the gap in the
quasiparticle spectrum.
Simultaneously the quasiparticle peaks broaden significantly above the critical
temperature, resembling the observed pseudogap behavior in
high-$T_c$ superconductors.
\end{abstract}

]

The pseudogap, or depletion of the single particle spectral weight
around the Fermi level, \cite{Timusk} is the most striking
demonstration that cuprate superconductors are not described by
the BCS scenario of superconductivity. The pseudogap opens in the
normal state as the temperature is lowered below a crossover
temperature $T^{\ast}$ and extends over a wide range of
temperatures in the underdoped cuprates. ARPES \cite{Campuzano,Feng}
and the scanning tunneling spectroscopy (STS) (see Refs. in
\cite{Timusk}) provide particularly complete information about the
pseudogap behavior and show a smooth crossover from the pseudo- to
superconducting (SC) gap. The transition from SC to normal
behavior appears to be driven by phase fluctuations \cite{Corson}
and  is well described by the Berezinskii-Kosterlitz-Thouless
(BKT) theory of vortex-pair unbinding.

There are currently many possible explanations for the unusual
behavior of HTSC. One of these is based on the nearly
antiferromagnetic Fermi liquid model \cite{Pines}. Another
explanation, proposed by Anderson, relies on the separation of the
spin and charge degrees of freedom.\cite{Anderson} The third
approach, which we follow in this paper, relates the observed
anomalies to precursor SC fluctuations, and in particular
fluctuations in the phase of the complex ordering field, as
originally suggested by Emery and Kivelson \cite{Emery}.

The phase diagram for a simple microscopic 2D model (see (\ref{Hamilton})
below) which formalizes the scenario of \cite{Emery} has been studied
in.\cite{Gusynin.JETP} The Green's function (GF) for this model was derived in
\cite{Gusynin.JETPLett} using the correlator
$\langle\exp(i\theta(x)/2)\exp(-i\theta(0)/2)\rangle$ for the phase
fluctuations in the classical (static) approximation and neglecting the
coupling between the spin and charge degrees  of freedom.

The associated spectral function (SF) $A(\omega,{\bf
k})=-(1/\pi){\rm Im}G(\omega+i0,{\bf k})$ has also been derived
analytically in \cite{Gusynin.JETPLett}. Being proportional to the
intensity of ARPES \cite{Campuzano}, the SF encodes information
about the quasiparticles and pseudogap. The result of
\cite{Gusynin.JETPLett} showed that, while the temperature
behavior of the quasiparticle peaks is in correspondence with
experiment \cite{Campuzano}, the gap in the spectrum remains
unfilled.\cite{Evans} In an earlier paper \cite{Franz} filling of
the gap has been achieved as a result of a Doppler shift in the
quasiparticle excitation spectrum. This Doppler shift originated
from the semiclassical  coupling of the mean field $d$-wave
quasiparticles to the supercurrents induced by classically
fluctuating unbound vortex-antivortex pairs. It was shown in
\cite{Kwon} that the shift can be identified with the coupling of
the spin and charge degrees of freedom.

The purpose of the present paper is to show analytically that
filling of the gap can also result  from the quantum
(dynamical) phase fluctuations, even when the coupling between
charge and spin degrees of freedom is entirely neglected. 
As pointed out in \cite{Tremblay.new} the mechanism of the gap
filling has to be understood yet even for the simplest attractive 
Hubbard model. Thus we contribute into the discussion \cite{Tremblay.new} 
as to which mechanism or mechanisms lead to pseudogap filling but within 
the scenario based on phase fluctuations, by studying the mechanism proposed
in \cite{Capezzali}. The advantage of our calculation is that 
the SF is obtained as a single integral with an analytical integrand, 
no numerical analytical continuation was necessary resulting in far 
greater accuracy.

Let us consider the continuum version of the 
2D attractive Hubbard model defined by the
Hamiltonian density:
\begin{equation}
\hspace{-2mm} {\cal H} = - \psi_{\sigma}^{\dagger}(x) \left(\frac{\nabla^{2}}{2
m} + \mu \right)
    \psi_{\sigma}(x) - V \psi_{\uparrow}^{\dagger}(x)
       \psi_{\downarrow}^{\dagger}(x) \psi_{\downarrow}(x)
       \psi_{\uparrow}(x),
\label{Hamilton}
\end{equation}
where $x= \mbox{\bf r}, \tau$ denotes the space and imaginary time
variables, $\psi_{\sigma}(x)$ is a fermion field with spin $\sigma
=\uparrow,\downarrow$, $m$ is the effective fermion mass, $\mu$ is
the chemical potential, and $V$ is an effective local attraction
constant; we take $\hbar = k_{B} = 1$. Clearly the Hamiltonian
(\ref{Hamilton}) is too simple to be adequate for systems as
complex as cuprate HTSC. However, it has proved itself as a very
convenient model for both numerical, in particular Monte Carlo
\cite{Ran1,Ran2} simulations, and analytical approaches
\cite{Gusynin.JETP,Gusynin.JETPLett,Capezzali,Babaev}
which does exhibit gap-like behavior above $T_c$
(see also Refs. in the review \cite{we.review}). Moreover
one may use the model to obtain a fully analytic treatment of the
pseudogap properties, and apply such results to obtain a better
understanding of more complex and less tractable models.

The calculation of the GF is performed in Nambu variables and the
fermions are
treated as composite objects, comprising both spin and charge parts:
\begin{equation}
\Psi^{\dagger}(x) = \left( \begin{array}{lr}
\psi_{\uparrow}^{\dagger}(x) &  \psi_{\downarrow}(x)
\end{array} \right) =
\Upsilon^{\dagger} (x) \exp[-i\tau_{3}\theta(x)/2],
\label{Nambu.spinor}
\end{equation}
where $\Upsilon^{\dagger}(x)$ is the Nambu spinor of neutral
fermions.  Substituting (\ref{Nambu.spinor}) into the standard
definition of the GF
$G(x) =\langle \Psi(x) \Psi^{\dagger} (0) \rangle$ one obtains
the GF of the charged (observed) fermions
\begin{equation}
G_{\alpha \beta}(x) = \sum_{\alpha^\prime, \beta^{\prime}} {\cal
G}_{\alpha^\prime\beta^\prime}(x)
\langle (e^{i\tau_{3}\frac{\theta(x)}{2}})_{\alpha\alpha^\prime}
(e^{-i\tau_{3}\frac{\theta(0)}{2}})_{\beta^\prime\beta} \rangle,
\label{Green.splited}
\end{equation}
as the product of the GF for the neutral fermions ${\cal
G}_{\alpha\beta}(x)=\langle \Upsilon_{\alpha}(x)
\Upsilon^{\dagger}_{\beta}(0) \rangle$ and the phase correlator
$\langle\exp(i\tau_{3} \theta(x)/2) \exp(-i
\tau_{3}\theta(0)/2)\rangle$. For the frequency-momentum
representation of Eq.~(\ref{Green.splited}) one has
\begin{eqnarray}
G(i \omega_{n}, \mbox{\bf k})& = & T \sum_{m = - \infty}^{\infty} \int
\frac{d^2 p}{(2 \pi)^{2}} \sum_{\alpha,\beta=\pm} P_\alpha {\cal G}(i
\omega_{m}, \mbox{\bf p}) P_\beta \nonumber \\
&& \times D_{\alpha\beta} (i \omega_{n} - i \omega_{m}, \mbox{\bf k} -
\mbox{\bf p}), \label{Green.projectors.momentum}
\end{eqnarray}
where $P_{\pm}={1\over2}(\hat I\pm \tau_3)$ are the projectors;
$\hat I$ and $\tau_{3}$ are unit and Pauli matrices;
$D_{\alpha\beta}(i\Omega_m, \mbox{\bf q})$ is the Fourier
transformation of the phase correlator $D_{\alpha\beta}(x) =
\langle\exp(i\alpha \theta(x)/2) \exp(-i
\beta\theta(0)/2)\rangle$; $\omega_{n} = (2n+1) \pi T$ and
$\Omega_{n} = 2n \pi T$ are respectively odd and even Matsubara
frequencies. The GF of neutral fermions is taken
in the mean-field approximation (see \cite{Gusynin.JETP,Gusynin.JETPLett})
\begin{equation}
{\cal G}(i \omega_{n}, \mbox{\bf p}) = - \frac{ i \omega_{n} \hat{I} +
\tau_{3} \xi(\mbox{\bf p}) - \tau_{1} \rho} {\omega_{n}^{2} +
\xi^{2}(\mbox{\bf p}) + \rho^{2}} \,,  \, \, \xi(\mbox{\bf p}) =
\frac{\mbox{\bf p}^{2}}{2m} - \mu \label{Green.neutral}
\end{equation}
with $\mbox{\bf p}$  being a 2D vector and
$\rho\equiv\langle\rho(x)\rangle$, where $\rho(x)$ is the modulus
of a complex ordering field $\Phi(x) = \rho(x) \exp [i\theta(x)]$.
Note that $\langle \Phi (x) \rangle = 0$ for $T \neq 0$ as it
should be in 2D in accordance with the
Coleman-Mermin-Wagner-Hohenberg theorem, while the value of $\rho$
is allowed to be nonzero. Note also that the GF
(\ref{Green.projectors.momentum}) does not contain the symmetry
violating term $\sim \tau_{1}$ which could originate from the
terms $P_{\pm} {\cal G}(i \omega_{n}, \mbox{\bf k}) P_{\mp}$ since
the correlators $D_{+-} = D_{-+} =0$ \cite{Gusynin.JETPLett}.

The representations (\ref{Green.splited}), (\ref{Green.projectors.momentum})
for the fermion GF with decoupled spin and charged degrees of freedom
are appropriate when one can neglect the fluctuations of the modulus
$\rho(x)$ and when the energy of the phase distortions
is smaller than the energy gain due to nontrivial $\rho$. 
For the present s-wave model this means that the condition $\rho \gg T$
should be satisfied, so that the modulus-phase representation
(\ref{Nambu.spinor}) should be useful even for $T$ close to $T_{\rm
BKT}$ and allows one to study the evolution of the SC gap to the
pseudogap. The region of temperatures $T$ where the condition
$\rho \gg T$ is satisfied depends crucially on the relative size
of the pseudogap region $(T^{\ast}- T_{\rm BKT})/T^{\ast}$ which
may be reasonably large in 2D for intermediate
coupling \cite{Tremblay}. The pseudogap in the current work is the
result of the low dimensionality of the system and does not need
the existence of preformed local pairs. The nonzero value of
$\rho$ is due to average local density of Cooper pairs which do
not have coherence above $T_{\rm BKT}$.

The representation (\ref{Green.projectors.momentum}) shows
(see the explanation after Eq.~(\ref{Green.neutral}))
that the GF for the charged fermions is defined by the correlator
$D(x) \equiv D_{++}(x) = D_{--}(x)$.
The asymptotic form of the correlator at large distances,
describing also the temporal decay of correlations, is given by
the form \cite{Capezzali,Huber}
\begin{equation}
D^{R}(t, \mbox{\bf r}) = \exp (- \gamma t)
(r/r_{0})^{-T/8 \pi J}
\exp ( - r/\xi_{+}(T)).
\label{correlator.dynamic}
\end{equation}
Here $t$ is the real time, $\gamma$ is a decay constant, $r_0
\equiv (2/T) (J/K)^{1/2}$ is the scale for the algebraic decay of
correlations in the SC BKT phase ($T < T_{\rm BKT}$, where $T_{\rm
BKT}$ is the temperature of the BKT transition), $\xi_{+}(T)$ is
the phase coherence length for $T > T_{\rm BKT}$ ($\xi_{+}(T \to
T_{\rm BKT}^{+}) \to \infty$). The constants $J$ and $K$ are the
bare (mean-field) superfluid stiffness and compressibility, respectively 
which have been calculated in \cite{Gusynin.JETP}.

Previously, using the representation (\ref{Nambu.spinor}), only
the classical (static $\Omega_n = \gamma =0$) fluctuations have
been considered analytically \cite{Gusynin.JETPLett}. The Fourier
transform of (\ref{correlator.dynamic}) for this case is
\begin{equation}
D(i \Omega_{n}, \mbox{\bf q}) \simeq \delta_{n, 0} C
[\mbox{\bf q}^2 + (1/\xi_{+})^{2} ]^{-\alpha} /T \,,
\label{Fourier.D.final}
\end{equation}
where $C \equiv 4 \pi (\Gamma(\alpha) / \Gamma(1-\alpha) )
( 2/r_{0} )^{2\alpha - 2}$ and $\alpha \equiv 1 - T / 16 \pi J$.
For $T \sim T_{\rm BKT}$ the value of
$\alpha \simeq 1 - T/32 T_{\rm BKT}$.
The presence of $\alpha \ne 1$ in (\ref{Fourier.D.final})
is related to the preexponent factor $(r/r_0)^{-T/8 \pi J}$ in 
(\ref{correlator.dynamic}). This factor was not included in the analysis 
of \cite{Franz,Kwon}, but the treatment of \cite{Capezzali} includes
this factor.  

One can now extend the analysis to the
case of quantum (dynamical) phase fluctuations.
We propose the following generalization of
(\ref{Fourier.D.final}):
\begin{equation}
D(i \Omega_{n}, \mbox{\bf q})  = \frac{C (v^2)^{\alpha}}
{T [v^2 \mbox{\bf q}^2 + (v/\xi_{+})^{2} + \Omega_{n}^{2} +
2 \gamma |\Omega_{n}|]^{\alpha}},
             \label{Fourier.D.dynamical}
\end{equation}
where $v$ is the velocity of the Bogolyubov excitations. Recall that in 2D $v
= v_{F}/\sqrt{2}$, where $v_{F}$ is the Fermi velocity.
The asymptotic form of the retarded GF corresponding to the GF
(\ref{Fourier.D.dynamical}) is
\begin{eqnarray}
& D^{R} (t, \mbox{\bf q}) \sim \nonumber \\
& \left\{
\begin{array}{c c}
t^{\alpha-1} e^{- \gamma t} \,, &
v^2 q^2 > \gamma^{2} - v^2 \xi_{+}^{-2} \\
t^{\alpha-1}
e^{-t (\gamma - \sqrt{\gamma^2 - v^2 \xi_{+}^{-2} - v^2 q^{2}})} \,, &
v^2 q^2 < \gamma^{2} - v^2 \xi_{+}^{-2}
\end{array} \right.  \nonumber \\
& \qquad \qquad  \qquad t \to +\infty \, . 
\label{D.retarded.dynamical}
\end{eqnarray}
Eq.~(\ref{Fourier.D.dynamical}) can be regarded as a convenient
(for analytical studies) 
generalization of the phenomenological
dependence (\ref{correlator.dynamic}) \cite{Capezzali} which for
nonzero $\gamma$ includes the decay of the phase correlations due
to the presence of free vortices above $T_{\rm BKT}$.
One can see that for  $v^2 q^2 < \gamma^{2} -
v^2 \xi_{+}^{-2}$ the decay rate is less than $\gamma$ and it is minimal for
$q=0$. This means that, for large distances, phase fluctuations do not feel
the pair vortices which have smaller size.

In the present work both $\gamma$ and $\xi_{+}$ are
phenomenological parameters which can be derived from the theory
of the BKT transition. Since in the SC BKT phase the vortices are
confined, one can state \cite{Huber} that there is a critical
slowing down of the phase fluctuations when the temperature
approaches $T_{\rm BKT}$, i.e. $\gamma(T\to T_{\rm BKT}^{+}) = 0$.
In fact the detailed theory of BKT transition predicts that 
$\gamma(T) \sim \xi_{+}^{-z}(T)$, where $z$ is the dynamical exponent
and  $\xi_{+}^{-1}(T) \simeq \xi_{0}^{-1}
\exp(- b/ \sqrt{T - T_{\rm BKT}})$, where $b$ is a positive
constant. In what follows we just restrict ourselves by comparing
two cases: $\gamma \neq 0$ above $T_{\rm BKT}$ and $\gamma =0$ for
$T < T_{\rm BKT}$.

We note that there is now experimental evidence
\cite{Corson} for the vortex-pair unbinding (BKT) nature of the SC
transition in the Bi-cuprates. We stress also that, despite the
rather simple form of the GF, it takes into account the presence
of vortices while the self-consistent $T$-matrix approximation
(see e.g. \cite{Tremblay.new,we.review,Levin}) cannot describe them.

The SF associated with GF (\ref{Green.projectors.momentum}) can readily be
expressed in terms of the corresponding SF for the GF (\ref{Green.neutral})
and (\ref{Fourier.D.dynamical}) as
\begin{eqnarray}
&& A(\omega, \mbox{\bf k}) =  - \frac{1}{\pi}{\rm Im} G_{11}(\omega + i0,
\mbox{\bf k}) = \int\limits_{-\infty}^\infty d\omega^\prime \left[\frac{1}{1 +
e^{\omega^\prime/T}} \right. - \nonumber  \\
&& \left.  \frac{1}{1 - e^{(\omega^\prime-\omega)/T}} \right]  \int\frac{d^2
p}{(2\pi)^{2}} a_F(\omega^\prime, \mbox{\bf p})a_B(\omega-\omega^\prime,
\mbox{\bf k} - \mbox{\bf p}) \,.
           \label{spectral.1}
\end{eqnarray}
where these SF are
\begin{eqnarray}
&& a_{F} (\omega, \mbox{\bf p}) = (\omega + \xi(\mbox{\bf p}))
\delta(\omega^{2} - \xi^{2}(\mbox{\bf p}) - \rho^2)  \mbox{sgn} \omega \,,
\nonumber               \\
&& a_{B}(\Omega, \mbox{\bf q}) = - \frac{1}{\pi} \mbox{Im}D^{R} (\Omega,
\mbox{\bf q}) \,.
            \label{spectral.input}
\end{eqnarray}
The $\delta$-function in $a_{F}$ allows one to perform the angular
integration
in (\ref{spectral.1}). Finally integrating over momentum the SF can
be expressed
as a single integral:
\begin{eqnarray}
&& A(\omega,{\bf k})= - \frac{1}{2 \pi^2 T}\frac{\Gamma(\alpha)}{\Gamma(1 -
\alpha)} \left( \frac{2}{m r_{0}^{2}} \right)^{\alpha-1} \int
\limits_{-\infty}^\infty d \omega^\prime {\rm sgn}(\omega^\prime + \omega)
\nonumber \\ && \times \left[ \frac{1}{1-e^{\omega^\prime/T}}-
\frac{1}{1+e^{(\omega^\prime+\omega)/T}}\right]
\frac{\theta[(\omega^\prime+\omega)^2- \rho^2]}
{\sqrt{(\omega^\prime+\omega)^2 - \rho^2}} \nonumber    \\
&& \times  \left\{ \left[ I(\omega,{\bf k},\omega^\prime)
\left(\omega^\prime+\omega+\sqrt{(\omega^\prime+\omega)^2- \rho^2}\right)
\right. \right. \nonumber \\ && \left. \times
\theta(\mu+\sqrt{(\omega^\prime+\omega)^2 -\rho^2}) \right]
\nonumber                 \\
&&\left.+\left[\sqrt{(\omega^\prime+\omega)^2-\rho^2}
\rightarrow -\sqrt{(\omega^\prime+\omega)^2-\rho^2}\right]\right\}\,,
                     \label{spectral.final}
\end{eqnarray}
where we have introduced a function $I(\omega,{\bf k},\omega^\prime)$:
\begin{eqnarray}
&& I(\omega,{\bf k},\omega^\prime)  = \nonumber \\
&&\pi{\rm Im} \left[(x_{-}-a- ib)^{-\alpha}
F\left(\frac{1}{2},\alpha;1;\frac{x_{+} -x_{-}}{a+ib-x_{- }}\right)\right]\,,
\nonumber \\ && a = \frac{1}{2m} \left(\frac{\omega^{\prime \: 2}}{v^2} -
\xi_{+}^{-2}\right)\,, \qquad b=\frac{\gamma\omega^\prime}{m v^2}\,, \nonumber \\
&& x_{\pm}=\left(\sqrt{\frac{k^2}{2m}}\pm
\sqrt{\mu+\sqrt{(\omega+\omega^\prime)^2-\rho^2}}\right)^2\,.
                     \label{I}
\end{eqnarray}

For $\gamma=0$ the expression (\ref{spectral.final}) can be
rewritten in the following form
\begin{eqnarray}
&& A(\omega,{\bf k})= - \frac{1}{2 \pi^2 T}\frac{\Gamma(\alpha)} {\Gamma(1-
\alpha)} \left( \frac{2}{m r_{0}^{2}}\right)^{\alpha-1}
\int\limits_{-\infty}^\infty d\omega^{\prime} {\rm sgn}\omega^{\prime}
\nonumber   \\
&& \times {\rm sgn}(\omega^{\prime}+ \omega)
\left[\frac{1}{1-e^{\omega^{\prime}/T}}-
\frac{1}{1+e^{(\omega^{\prime}+\omega)/T}}\right] \nonumber  \\
&& \times  \frac{\theta[(\omega^{\prime}+\omega)^2-\rho^2]}
{\sqrt{(\omega^{\prime}+\omega)^2-\rho^2}} \theta(\omega^{\prime
2}-v^2\xi^{-2}_{+}) \nonumber \\
&& \times \left\{\left[\pi(a-x_{-})^{-\alpha}
F\left({1\over2},\alpha;1;\frac{x_{+}-
x_{-}}{a-x_{-}}\right)\theta(a-x_{+})\right.\right.
\nonumber                \\
&& + (x_{+}-x_{-})^{-1/2}(a-x_{-})^{1/2-\alpha}B({1\over2}, 1-\alpha)
\nonumber \\ &&  \left. \times
F\left({1\over2},{1\over2};{3\over2}-\alpha;\frac{a-x_{-}}{x_{+}-x_{- }}
\right)\theta(x_{+}-a)
\theta(a-x_{- })\right] \nonumber \\
&& \times \left(\omega^{\prime}+\omega+\sqrt{(\omega^{\prime}
+\omega)^2-\rho^2}\right)\theta(\mu+\sqrt{(\omega^{\prime}+\omega)^2 -\rho^2})
\nonumber\\
&&\left.+\left(\sqrt{(\omega^{\prime}+\omega)^2-\rho^2}\rightarrow-
\sqrt{(\omega^{\prime}+
\omega)^2-\rho^2}\right)\right\} \label{zero.gamma}
\end{eqnarray}
which describes the case of the non-damped dynamical phase
fluctuations. While Eq.~(\ref{zero.gamma}) appears to be more
complicated than the more general Eq.~(\ref{spectral.final}), it
proves useful for making  a general statement about the gap
filling.

Expressions (\ref{spectral.final}) and (\ref{zero.gamma}) present
the main result of the paper for the SF in the case of quantum
phase fluctuations. Before proceeding to the discussion of their
numerical integration, we recap briefly the case of classical
phase fluctuations. 
The static SF, obtained in \cite{Gusynin.JETPLett}, is reproduced from 
(\ref{zero.gamma}) after changing the variable 
$\omega^\prime \to v \omega^\prime$ and taking formally the limit $v\to0$.
It was discussed in detail in \cite{Gusynin.JETPLett} and we would
like to stress here only two main points. The first is that the SF
is identically zero inside  the gap ($A_{cl}(\omega, \mbox{\bf k})
=0$ for $|\omega| < \rho$) \cite{Evans,foot1}. The second is that
in addition to the usual quasiparticle peaks it reveals extra
peaks at $\omega = \pm \rho$.

\begin{figure}
\centering{
\includegraphics[width=2.5in]{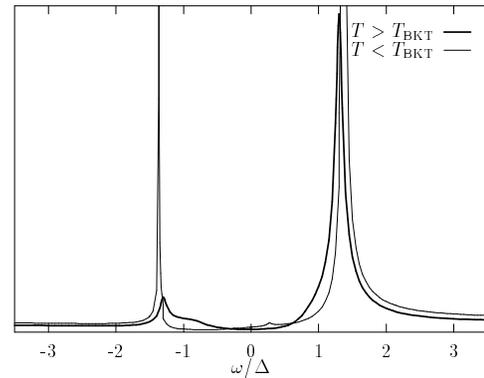}}
\caption{The spectral function $A(\omega, \mbox{\bf k})$ as a
function of $\omega$ in units of the zero temperature SC gap
$\Delta$ for $k > k_{F}$ taken at two different temperatures,
$T_{1} < T_{\rm BKT} < T_{2}$.} \label{fig:1}
\end{figure}

Let us now return to the discussion of the results of numerical
integration based on Eqs.~(\ref{spectral.final}) and
(\ref{zero.gamma}) which are shown in Fig.~\ref{fig:1}.

\noindent a) For $T < T_{\rm BKT}$ there are two highly pronounced
quasiparticle peaks at $\omega = \pm E(\mbox{\bf k}) \equiv \pm
\sqrt{\xi^{2}(\mbox{\bf k}) + \rho^{2}}$. Since below $T_{\rm
BKT}$ both $\xi_{+}^{-1}$ and $\gamma$ are zero, the width of the
peaks is almost entirely controlled by $\alpha$ whose deviation
from 1 gives the non-Fermi liquid behavior of the GF
\cite{Gusynin.JETPLett}. Since the value of $\alpha$
is very close to 1 this non-Fermi liquid behavior is hardly
distinguishable from ordinary widening of the quasiparticle peaks
due to damping. Furthermore, because $\alpha \to 1$ as $T \to 0$
the width of the peaks is temperature
dependent so that the peaks become sharper as $T$ decreases resembling
the data of ARPES for the anti-node direction.\cite{Campuzano}

We note also that $k \ne k_{F}$ in Fig.~\ref{fig:1} so that the 
quasiparticle peaks are not equal to each other in contrast to the case
of symmetrized ARPES data \cite{Timusk}. The reason for choosing
$k \ne k_{F}$ in Fig.~\ref{fig:1} was to prove that
the extra peaks present in the SF calculated for the classical phase 
fluctuations \cite{Gusynin.JETPLett,we.review} are now absent.

The gap in the SF remains almost unfilled and has ``U''-like shape.
We stress, however, that in contrast
to the static case \cite{Gusynin.JETPLett} the SF
(\ref{spectral.final}) is nonzero even for $|\omega| < \rho$ and
this is evidently related to the presence of the quantum
(dynamical) phase fluctuations, i.e. the terms with $\Omega_n \neq
0$. In particular, using Eq.~(\ref{zero.gamma}) one can check
analytically that even for $\gamma =0$ that due to these terms
$A(\omega = 0, {\bf k} ={\bf k}_{F}) \neq 0$.
We estimated the temperature $T_{cl}$ of the quantum to classical crossover 
in the way similar to that of \cite{Randeria}.
For the low carrier densities this gives
$T_{cl} \sim T_{\rm BKT}$, while for the high carrier densities, 
where the pseudogap phase shrinks \cite{Gusynin.JETP} 
we obtain $T_{cl} \sim \rho(T=0) \sim T^{\ast}$.
Thus, the crossover temperature is always too large (in the absence of
dissipation) for classical fluctuations to play a significant role at
low $T$.

\noindent b) For $T > T_{\rm BKT}$  the quasiparticle peaks at
$\omega \approx \pm E(\mbox{\bf k})$ are far less prominent. This
is caused by the fact that $\xi_{+}$ is now nonzero due to the
presence of vortices. Increasing the temperature further decreases
the value of $\xi_{+}$ so that the quasiparticle peaks become even
less pronounced. This behavior of the quasiparticle peaks
reproduces qualitatively the ARPES studies \cite{Campuzano,Feng} 
for the anti-node direction which
show clearly that the quasiparticle SF broadens dramatically when
passing from the SC to normal state. The width of the
quasiparticle peaks is controlled primarily by $\xi_{+}$ above
$T_{\rm BKT}$ since this width remains practically constant as
$\gamma$ changes. Since $\xi_{+}$ is the phase coherence length,
the current model supports the premise \cite{Feng} that the
quasiparticle peaks are related to the phase coherence, not to 
the energy gap. The deviation of $\alpha$ from 1 becomes, however,
sizable at $T \geq T_{BKT}$.

\noindent
c) As discussed above, for $T > T_{\rm BKT}$ the value
of $\gamma$ is nonzero and increases with increasing temperature.
This increase of $\gamma$ (typically up to values of the order
of $0.5 \mu$), together with the decrease of $\xi_+$, causes
filling of the gap and changes
its shape from ``U''-like to ``V''-like.  In other words,
the quasiparticle peaks grow ``shoulders''
which eventually fill the gap.  
We stress, however,  that the increase of $\gamma$  in fact diminishes 
the effect of the quantum phase fluctuations pushing
$T_{cl}$ down \cite{Randeria} which is seen from the 
reducing of $A(\omega =0, {\bf k} ={\bf k}_F)$ as $\gamma$ grows 
(see Fig.~1, where the bottom parts of the curves cross each other),
but the abovementioned ``shoulders'' simultaneously emerge. 
Thus the filling of the gap at $T>T_{\rm BKT}$ in the present model is
caused by quantum phase fluctuations {\it in the presence} of nonzero 
dissipation because the damping $\gamma$ contributes only when
$\Omega_n\neq 0$.
   
\noindent d) Due to the smooth dependence of $\xi_{+}^{-1}$ and $\gamma$ on
$T$ as the temperature passes through $T_{\rm BKT}$ there is no sharp
transition at $T_{\rm BKT}$ in agreement with
experiment \cite{Timusk,Campuzano}.
A  similar filling of the gap was obtained in \cite{Capezzali} for
$\gamma = 0.5 \mu$ where the correlator (\ref{correlator.dynamic}) was used
for the numerical computation of the self-energy of the fermions and the
subsequent extraction of the SF from the fermion GF. A recent Monte Carlo
simulation \cite{Tremblay} also shows similar behavior for the quasiparticle
peaks and the filling of the gap.  However it is the analytical character of
the present work, which relies on the explicit introduction of the charge and
spin degrees of freedom (\ref{Nambu.spinor}) for the Nambu spinors, that
enables us to unambiguously state the correspondence between the parameters of
the model and the observed features of the SF (\ref{spectral.final}).  

In spite of some similarities between the results observed and
those just obtained, the latter are only illustrative since we have
considered a model with non-retarded $s$-wave attraction. 
However, it is likely that for
$d$-wave pairing, the properties obtained can be used for the
description of the systems in the anti-node direction on the Fermi
surface. 
It is however essential to consider fluctuations in the
modulus of the order parameter to extend the analysis to the nodal
directions on the Fermi surface. 
The value of $\gamma = 0.5 \mu$
which results in a filled gap appears to be too large to be due to
the vortex-vortex interaction. This leads one to the conclusion
that the mechanism considered here for gap filling may well not be
the only possible mechanism and that other interactions which lead
to the filling of the gap above $T_c$, see \cite{Franz,Kwon,Tremblay.new}, 
are also important.

In summary we studied the effect of the fluctuations
in the phase of the complex ordering field on the properties of a
2D system with four-fermion attraction. The fermion SF has been
given as a single integral over a function known in closed form.
Through the use of analytical techniques, we have been able to
demonstrate that the quantum phase fluctuations in the presence of
dissipation lead to the filling of the pseudogap even if one 
ignores the spin-charge coupling.

S.G.Sh  thanks Prof.~H.~Beck and Dr.~M.~Capezzali for stimulating
discussion.
This work was partly supported by the NRF, South Africa (R.M.Q and S.G.Sh),
by SCOPES-project 7UKPJ062150.00/1 (V.P.G., V.M.L. and S.G.Sh) and
by the research project 2000-061901.00/1  (S.G.Sh)
of the Swiss National Science Foundation.

\end{document}